\documentclass[aps,prl,floatfix,twocolumn,superscriptaddress]{revtex4-2}
\usepackage{color}

\usepackage{graphicx,
            amsmath,
            braket,
            xcolor,
            xspace,
            lipsum}
\begin{document}

\title{Mid-circuit qubit measurement and rearrangement in a $^{171}$Yb atomic array}

\author{M.~A.~Norcia}
\email[Email: ]{matt@atom-computing.com}
\author{W.~B.~Cairncross}
\author{K.~Barnes}
\author{P.~Battaglino}
\author{A.~Brown}
\author{M.~O.~Brown}
\author{K.~Cassella}
\author{C.-A.~Chen}
\author{R.~Coxe}
\author{D.~Crow}
\author{J.~Epstein}
\author{C.~Griger}
\author{A.~M.~W.~Jones}
\author{H.~Kim}
\author{J.~M.~Kindem}
\author{J.~King}
\author{S.~S.~Kondov}
\author{K.~Kotru}
\author{J.~Lauigan}
\author{M.~Li}
\author{M.~Lu}
\author{E.~Megidish}
\author{J.~Marjanovic}
\author{M.~McDonald}
\author{T.~Mittiga}
\author{J.~A.~Muniz}
\author{S.~Narayanaswami}
\author{C.~Nishiguchi}
\author{R.~Notermans}
\author{T.~Paule}
\author{K.~A.~Pawlak}
\author{L.~S.~Peng}
\author{A.~Ryou}
\author{A.~Smull}
\author{D.~Stack}
\author{M.~Stone}
\author{A.~Sucich}
\author{M.~Urbanek}
\author{R.~J.~M.~van de Veerdonk}
\author{Z.~Vendeiro}
\author{T.~Wilkason}
\author{T.-Y.~Wu}
\author{X.~Xie}
\author{X.~Zhang}
\author{B.~J.~Bloom\\ Atom Computing, Inc.}
\email[Email: ]{bbloom@atom-computing.com}

\begin{abstract}
\noindent Measurement-based quantum error correction relies on the ability to determine the state of a subset of qubits (ancillae) within a processor without revealing or disturbing the state of the remaining qubits.  Among neutral-atom based platforms, a scalable, high-fidelity approach to mid-circuit measurement that retains the ancilla qubits in a state suitable for future operations has not yet been demonstrated.  In this work, we perform imaging using a narrow-linewidth transition in an array of tweezer-confined $^{171}$Yb atoms to demonstrate nondestructive state-selective and site-selective detection.  By applying site-specific light shifts, selected atoms within the array can be hidden from imaging light, which allows a subset of qubits to be measured while causing only percent-level errors on the remaining qubits.  As a proof-of-principle demonstration of conditional operations based on the results of the mid-circuit measurements, and of our ability to reuse ancilla qubits, we perform conditional refilling of ancilla sites to correct for occasional atom loss, while maintaining the coherence of data qubits.  Looking towards true continuous operation, we demonstrate loading of a magneto-optical trap with a minimal degree of qubit decoherence.  
\end{abstract}
\maketitle

\begin{figure}[!htb]
		\includegraphics[width=3.38 in]{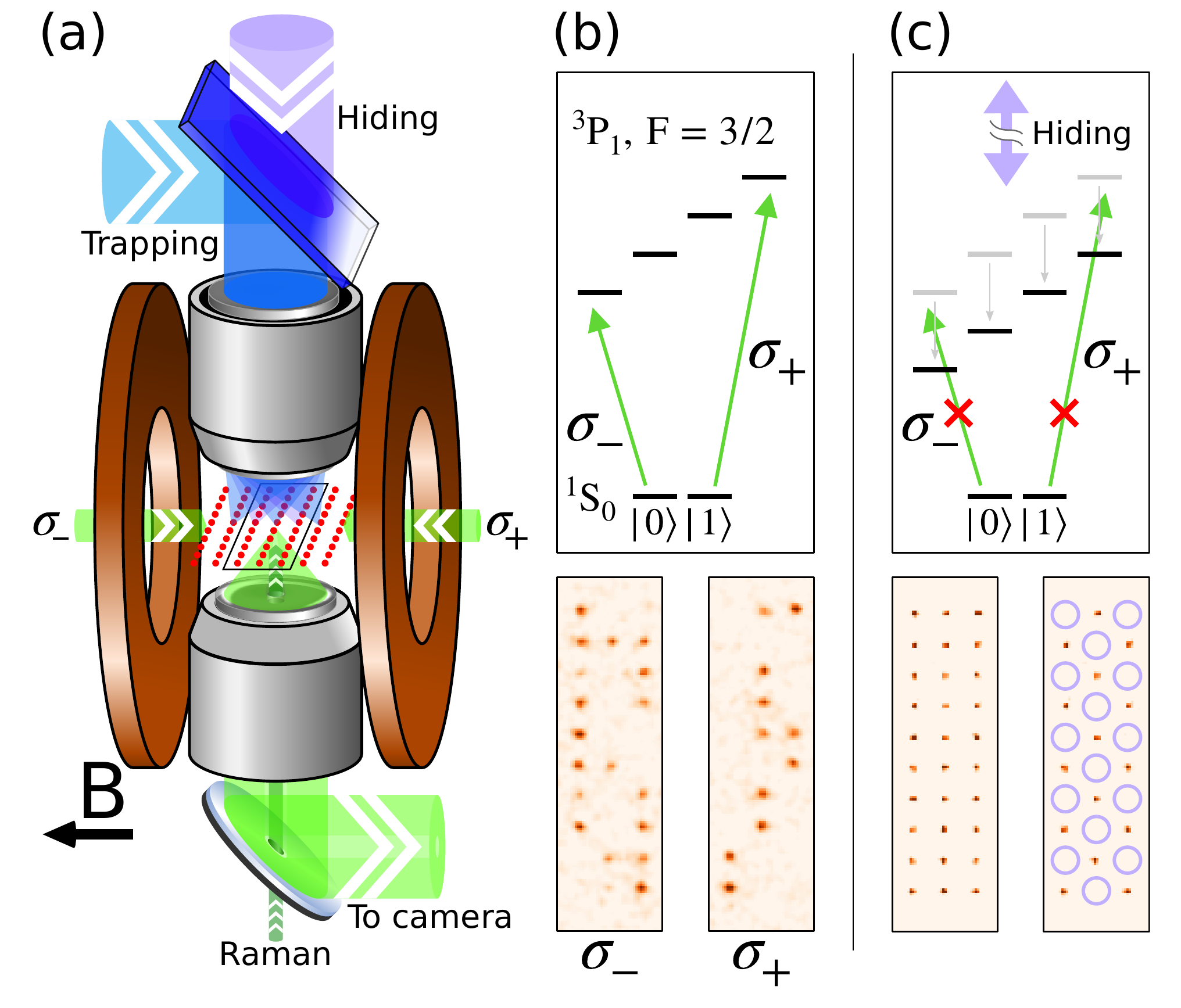}
		\caption{\textbf{(a)} Experimental diagram.  Individual $^{171}$Yb atoms are trapped in the sites of an optical tweezer array in the presence of a 500~Gauss magnetic field.  Two high-numerical-aperture objectives allow for site-resolved imaging, and the targeted application of trapping light (483~nm wavelength) and hiding light (460~nm wavelength).  A movable tweezer (483~nm wavelength) rearranges atoms between trapping sites.  A 556~nm laser incident through a hole in the imaging objective is used to drive global Raman transitions between the qubit states.  \textbf{(b)} Laser beams with opposite circular polarization and different frequencies are applied along the direction of the magnetic field to selectively image the two qubit states $^1$S$_0$ $m_f = 1/2, -1/2$ (which we label $\ket{1}$, $\ket{0}$) by coupling them to either of $^3$P$_1$ $m_f = 3/2, -3/2$ (556~nm wavelength).  The lower panels show two subsequent single-shot images of a fully filled 10 by 3 array of atoms prepared in an equal superposition of $\ket{1}$ and $\ket{0}$ prior to the first image. This shows projection of the qubit state and state-resolved, nondestructive imaging.  
        \textbf{(c)} Hiding light (purple arrows) applies up to 74~MHz differential light-shift between the $^1$S$_0$ and $^3$P$_1$ manifolds of targeted sites, pushing the imaging light off resonance.  The lower panel shows images of atoms prepared in $\ket{1}$ averaged over many trials without hiding light (left), and with hiding light applied to a checkerboard pattern of sites within the array (right). Purple circles indicate hidden sites.  }
		\label{fig:fig1}
\end{figure}

Useful error-corrected quantum computers must remove entropy from the processor faster than it can enter.  At the physical qubit level, entropy enters through errors due either to the interaction of the qubit with its environment or to imperfect control of a qubit. To achieve protection from these errors, entropy must be transferred to and removed from a subset of qubits (``ancilla qubits") without inducing undue errors on other ``data" qubits.  Measurement-based quantum error correction --- which involves repeated ``mid-circuit" measurement of the ancilla qubits --- represents a promising approach to entropy removal \cite{shor1995scheme, knill1997theory, dennis2002topological, fowler2012surface, Egan2021faulttolerant, ryananderson2022implementing, Acharya2023suppressing}.  

In quantum processors based on individually controlled atoms (whether neutral or ionized), measurement is typically performed by collecting light resonantly scattered by the atoms from a laser beam.  Because atoms of the same species generally share the same resonance frequencies, and because many photons must typically be scattered from each atom to accurately determine its quantum state, performing measurements on a subset of atoms without introducing errors on others is difficult.  Beyond the challenges of directing laser light solely on a subset of closely spaced atoms, photons scattered by the atoms being measured could be reabsorbed by others, and even a single scattering or absorption event can be enough to problematically modify the quantum state of a data qubit.  

One approach to mitigate this issue is to use different atomic species (which can have very different resonance frequencies) for data and ancilla qubits.  This technique is now commonly used for trapped-ion based processors \cite{Erickson2022highfidelity}, and has recently been explored for neutral-atom based systems \cite{singh2022mid}.  Unfortunately, the use of multiple species adds significant technical complexity and relies on inter-species gates.  These have been successfully implemented using ions \cite{Hughes2020Benchmarking}, but have yet to be demonstrated for neutral atoms.  Scattering of imaging light from data qubits can also be reduced by using spatially separated regions for readout \cite{kielpinski2002architecture, Pino2021Demonstration, bluvstein2022quantum}, though this has yet to be demonstrated for free-space detection of individual neutral atoms.  Alternatively, mid-circuit measurements have been demonstrated with neutral atoms by selectively transferring from the qubit manifold into states that have very different coupling to the imaging light \cite{graham2023mid}, though this approach suffers from limited fidelity associated with the complex pulse sequences required to transfer the atomic state and with the required use of non-optimal imaging and cooling parameters.  Finally, within neutral-atom based systems, mid-circuit measurements have been performed by coupling atoms to high-finesse optical cavities \cite{Deist2022mid}, which reduces both the number of photons that must be scattered to resolve the state of the atom as well as the spatial distribution of scattered photons.  However, a scalable (parallelizable) cavity-based approach has yet to be demonstrated.

In this work, we present a scalable and high-fidelity method for mid-circuit measurement in a single-species tweezer-array of neutral $^{171}$Yb atoms (Fig.~\ref{fig:fig1}a).  Tweezer-confined alkaline earth atoms \cite{cooper2018alkaline, norcia2018microscopic, saskin2019narrow, barnes2022assembly}, particularly $^{171}$Yb \cite{jenkins2022ytterbium, ma2022universal} have recently emerged as a promising system for quantum computation.  Our approach is based on imaging light scattered from a narrow-linewidth transition \cite{saskin2019narrow, urech2022narrow}, which, when combined with Zeeman shifts and light shifts, allows us to perform highly state-selective and site-selective imaging (Fig.~\ref{fig:fig1}b,c).  This enables high-fidelity readout of arbitrarily chosen ancilla qubits within the array, while imparting only percent-level loss of contrast on non-measured qubits.  Prior demonstrations of mid-circuit measurement in neutral atoms based on free-space detection do not retain the ancilla in a suitable condition for future use, either because state-selectivity is achieved by inducing state-selective loss (requiring re-loading) \cite{singh2022mid}, or because heating during the mid-circuit measurement prevented resetting of the ancilla qubit state \cite{graham2023mid}.  In contrast, our approach has a high probability of retaining the ancilla qubits in a suitable condition for further use.  As a demonstration of how our technique can be used for conditional branching to correct for errors, as well as our ability to perform reset and repeated measurements of the same ancilla qubits, we perform optical pumping and repeated cycles of imaging on the ancilla qubits, while maintaining the coherence of data qubits.  Between these repeated measurements, we conditionally refill ancilla sites in the rare cases where they undergo atom loss.  Looking towards continuous operation beyond the lifetime of an individual atom within a tweezer, we show that a magneto-optical trap can be loaded while maintaining coherence among our data qubits.

\section{state-resolved nondestructive measurement}\label{stateresolved}

The ability to determine the quantum state of an atom without it being lost from the trap is a useful capability for mid-circuit measurements.  In many neutral-atom systems, the state of single atoms is determined by introducing state-selective loss followed by state-independent imaging of the remaining atoms \cite{nelson2007imaging, cheuk2016observation, parsons2016site, singh2022mid}.  Such approaches cannot distinguish atoms that populate the ejected state from those that already underwent loss, and would require frequent reloading and state preparation of new ancilla qubits if used in an error-correction protocol.  Non-destructive approaches to state-resolved measurement for single neutral atoms include physically separating qubit states \cite{boll2016spin, wu2019stern, Koepsell2020robust}, taking advantage of very different scattering rates between two states \cite{fuhrmanek2011free, gibbons2011nondestructive, martinez2017fast, kwon2017parallel} including through shelving in metastable optically excited states \cite{norcia2019seconds, madjarov2019atomic, barnes2022assembly}, or using high-finesse optical cavities \cite{bochmann2010lossless, gehr2010cavity, eto2011projective, Deist2022mid}.

We employ a simple alternative method that combines narrow-line imaging with large Zeeman shifts created by operating in a 500~Gauss magnetic field.  We selectively image the $^1$S$_0$ $m_f = 1/2, -1/2$ qubit states, which we label $\ket{1}$, $\ket{0}$, respectively, by applying light from either one of two counter-propagating imaging beams, which are each tuned to address one of the $^3$P$_1$ $m_f = \pm 3/2$ states. This provides access to narrow linewidth ($\sim 180$~kHz) closed cycling transitions.  Scattering from the $m_f = \pm 1/2$ excited states, which would allow population leakage between the qubit states, is suppressed by the large ratio of Zeeman shifts (771~MHz, 681~MHz between the $-3/2$ and $-1/2$ states and the $1/2$ and $3/2$ states, respectively) to transition linewidth.  Our 500~G magnetic field is selected to minimize such unwanted scattering, as well as other potential errors (described in Appendix C), while being consistent with mechanical and thermal constraints.  
Scattered light is collected by a high-numerical-aperture objective and imaged onto a low-noise camera. For each site in the tweezer array, we apply a threshold to the counts in an integration region to determine if an atom in the imaged state occupies that site.  We typically operate with an imaging duration of 5~ms, during which we register approximately 30 photons from a bright atom, which is chosen to balance distinguishability, loss, data-rate, and robustness to experimental drifts.

\begin{table}[h!]
\centering
\begin{tabular}{|p{1.2cm} | p{1cm} | p{1.3cm}| p{1.3cm}|p{1.3cm}|p{1.3cm}|} 
 \hline
 Prepare & Image & Process & 
 \multicolumn{3}{c|}{Error Per Image} \\
 \cline{4-6}

& & & Base & Data & Ancilla \\
 
 \hline \hline

 mixture & $\ket{1}$ & overlap & .001(1)& NA &.001(1)\\ 
 mixture & $\ket{0}$ & & .002(1) & NA & .001(1) \\

 \hline
 $\ket{1}$ & $\ket{1}$ & loss &  .005(2)  &.002(2) & .007(3) \\ 
\cline{3-6}
&  & $\ket{1}$ $\rightarrow$ $\ket{0}$ &  .0001(2)  &.0001(2) & .003(1) \\ 
\cline{2-6}

& $\ket{0}$ & loss & .0007(3)&.001(2) &  .003(2) \\ 
\cline{3-6}
  &  & $\ket{1}$ $\rightarrow$ $\ket{0}$ & .0001(2) & .000(1) & .000(1)\\

\hline
 $\ket{0}$ & $\ket{0}$ & loss & .010(2) & .000(1) & .020(5) \\ 
\cline{3-6}
  &  & $\ket{0}$ $\rightarrow$ $\ket{1}$ & .0003(1) &.0001(2) & .0001(2)\\ 
\cline{2-6}

  & $\ket{1}$ & loss & .0004(2) &.0009(3) & .0012(3) \\ 
\cline{3-6}
  &  & $\ket{0}$ $\rightarrow$ $\ket{1}$ & .0001(2) &.000(1) & .001(2)\\ 

 \hline
\end{tabular}
\caption{Imaging errors for base condition (no hiding light applied), and for data and ancilla qubits with hiding light applied to data qubits.  Uncertainties represent a Wilson score interval, where we report only the larger direction for visual clarity.  These results are averaged over a 10 by 7 site array.  }
\label{table:1}
\end{table}

Achieving low off-resonant scattering while maintaining cooling from the imaging beam is enabled by operating with low imaging beam power and relatively small red detuning (of order the transition linewidth).  This is facilitated by using ``magic" traps, where ground and excited states experience the same trapping potential.  We find a magic wavelength for the $^1$S$_0$ to $^3$P$_1$ $m_f = \pm 3/2$ transitions near 483~nm. The trapping light is polarized perpendicular to the magnetic field.  Magic traps also ease array uniformity requirements.

We characterize our imaging in terms of the accuracy with which we can distinguish the state of the atom, the probability of leakage into the other qubit state during imaging, and the probability of atom loss (Table \ref{table:1}).  
To measure loss from the imaged state, we perform repeated images and fit an exponential decay to the measured occupancy (apparent loss from the imaged state includes atom loss from the trap, as well as transitions to other unmeasured states, including the other qubit state).  We estimate distinguishability based on the overlap of a double-Gaussian fit to the count histograms obtained from stochastically occupied sites.  
To characterize rates of population leakage during imaging (which are much smaller than loss rates), we prepare atoms in either qubit state, perform a pre-image of that state (for these measurements, the array is stochastically loaded, so this pre-image allows us to post-select on occupied sites), then perform a ``dummy image" of each state, which uses our default imaging intensities and detunings     but a longer duration chosen to enhance our signal relative to statistical or systematic readout errors.  Finally, we image the population of either qubit state to infer population transfer.  Our reported values are inferred by dividing the measured population transfer by the ratio of the dummy image length to our default value of 5~ms.  Further details on our measurement procedure are provided in Appendix B.  

The largest error channel is loss of the imaged state.  For a range of sufficiently low scattering rates we observe a regime where the probability of loss is proportional to the number of collected photons --- typically near $10^{-4}$ per photon collected, which we estimate corresponds to approximately $4 \times 10^{-6}$ per photon scattered.  About half of this loss may be explained by the predicted Raman scattering out of $^3$P$_1$ due to the trap light in our approximately 350~$\mu$K deep traps (Appendix C).  In principle, this loss mechanism could be eliminated by operating in shallower traps (and imaging more slowly), or by repumping from the metastable $^3$P$_0$ and $^3$P$_2$ states at the expense of state-selectivity.  Additional loss in this regime may be caused by photo-ionization \cite{ma2023high}.  For higher scattering rates (corresponding to several scattering events per trap period), the loss probability increases sharply, which we attribute to heating of the atoms.   We choose an operating condition with a scattering rate just below the onset of this additional loss.

\section{site-resolved coherence-preserving measurement}\label{siteresolved}
\begin{figure*}[htb]
		\includegraphics[width=2\columnwidth]{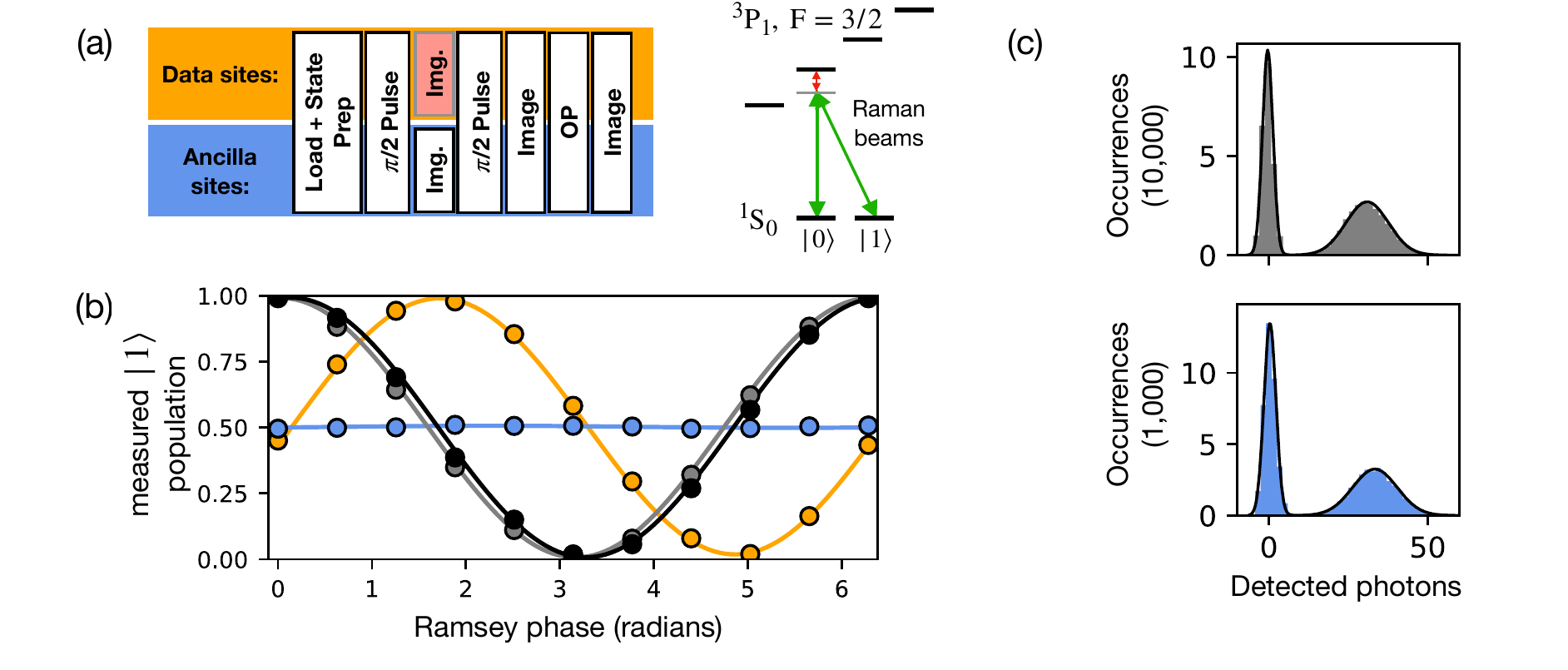}
		\caption{  Coherence-preserving mid-circuit measurement of sub-array.  \textbf{(a)} Experimental sequence for characterizing mid-circuit measurement, with operations performed on data sites indicated in the upper row and operations performed on ancilla sites in the lower row.  Measurements are performed on ancilla sites during a Ramsey sequence.  Data sites are shifted off resonance from imaging light, as indicated by red shading.  Optical pumping (OP) allows for post-selection of occupied sites based on a final image.  Coherent global Raman transitions between the two qubit states are driven by a laser 280~MHz detuned from $^3$P$_1$ $m_f = -1/2$ (red arrow), with two tones separated by the qubit frequency (388.9~kHz).  \textbf{(b)} Ramsey fringes (see main text for details) for no imaging or hiding light applied (black), hiding light but not imaging light applied (grey), and both imaging and hiding light applied to the array for data (orange) and ancilla (blue) sites.  (Hiding light is applied to alternating sites of the array, and defines the data sites.) Imaging of ancilla qubits causes a 1.3(8)\% loss of array-averaged contrast on data qubits, or 0.7(6)\% average single-site contrast loss (See Appendix D) as well as a $\Phi = $1.59(2) radian phase shift relative to the absence of imaging or hiding light.  Unless otherwise stated, all quantities presented in this work are ensemble averages, and uncertainties are smaller than the markers when not visible.  \textbf{(c)}  Imaging count histograms (relative to the average camera background) for stochastically occupied sites with no hiding light (upper) and for the mid-circuit measurement of ancilla sites from the dataset of part b (lower), showing minimal perturbation of the ancilla imaging when hiding data qubits.  The fitted peak separations are 31.3 and 33.0 photons, respectively.  For the data shown, a threshold of 9 signal photons enables a discrimination infidelity of 0.015\% (0.017\%) for the left (right) histograms, based on the overlap of a double-Gaussian fit.		}
		\label{fig:fig2}
\end{figure*}

In addition to enabling state-selective imaging, narrow-line imaging provides a means of site-selectivity.  By applying an additional tweezer array that addresses a subset of sites, at a wavelength with large differential polarizability between the qubit and $^3$P$_1$ states, we can hide selected qubits from the imaging light.  A similar approach has been previously demonstrated for site-selective hiding of strontium atoms from narrow-line imaging \cite{urech2022narrow}, though not in a regime where hidden atoms retain coherence.  
For this demonstration, we work with a checkerboard pattern of hiding spots, with data and ancilla qubits occupying complimentary sub-arrays of a 10 by 7 site, stochastically loaded array.
The hiding light is derived from a laser with 459.5960(5)~nm wavelength, a previously underexplored magic wavelength for the $^1$S$_0$ to $^3$P$_0$ transition  (Appendix F).  This wavelength is relatively near-detuned ($\sim 2$~nm) from the $^3$P$_1$ to $6s6d\,^3$D$_1$ transition, leading to a $\sim 9\times$ larger light shift of the $^3$P$_1$ state than the $^1$S$_0$ qubit states.  By using near-detuned light and primarily shifting a state that is unpopulated, it is possible to achieve large shifts with moderate laser powers, while causing relatively little unwanted scattering \cite{barnes2022assembly, burgers2022controlling}.  In our default operating conditions, the hiding-light shifts the imaging transition by $\Delta_h$=74~MHz, with a standard deviation of 6~MHz across sites in the array.

Our goal is to maintain coherence in the data sites with minimal change to the imaging quality on the ancilla sites.  In order to show that coherence is maintained in the data qubits, we embed the mid-circuit measurement in a Ramsey sequence (Fig.~\ref{fig:fig2}a).  We drive Raman transitions between the qubit states using a single laser beam with two tones, offset from each other by the qubit frequency and from the intermediate state ($^3$P$_1$ $m_f = -1/2$) by 280 MHz.  We typically operate with a two-photon Rabi frequency of 1.4~kHz.  The Raman beam contains polarization components both parallel to and perpendicular to the magnetic field, and so contributes both legs of the Raman transition \cite{jenkins2022ytterbium}.  Because of the large Zeeman splitting in $^3$P$_1$, the effect of coupling through the $m_f = 1/2$ state is suppressed.  We measure a Ramsey fringe (from which we can extract changes in the qubit coherence and phase) by performing two $\pi/2$ pulses using the Raman beams, and varying the phase of the second pulse by digitally scanning the differential phase of the tones sent into an acousto-optic modulator used to generate the two Raman beams.  We record the probability of obtaining a bright image of state $\ket{1}$ immediately after the second pulse, and perform post-selection using a final image to restrict our analysis to occupied sites within the stochastically-loaded array.
Without applying a mid-circuit measurement, the contrast of our Ramsey fringes is 98.5(6)\%, primarily limited by scattering from the intermediate state (this scattering could be reduced by using a larger intermediate state detuning and higher laser intensity).  Our ensemble-averaged qubit coherence time is 430(60)~ms, primarily limited by a magnetic field gradient across the array.  At our 500~Gauss operating field, the magnetic sensitivites of the $^3$P$_1$, $m_f = -3/2, -1/2, 1/2, 3/2$ states are -2.099, -0.550, 0.875, 2.099 MHz/Gauss respectively, and the magnetic sensitivities of the $^1$S$_0$, $m_f = \pm 1/2$ qubit states are $\mp 376.3$Hz/Gauss.  
A single echo pulse brings the coherence time to 3.3(2) seconds.  

When applying a mid-circuit measurement of $\ket{1}$ during the Ramsey hold time, we observe a contrast of 97.2(6)\%, representing a reduction of 1.3(8)\%, and a phase shift of $\Phi = $1.59(2) radians in the Ramsey fringe.  The phase shift is due to the differential coupling of the qubit states to $^3$P$_1$ via the imaging light, which is off-resonance in the presence of hiding light (Fig.~\ref{fig:fig2}b).  We attribute the loss of contrast primarily to a combination of residual off-resonant scattering of the imaging light, and the spread in phase shifts between sites (See Appendices C, D for details).  When fitting the Ramsey fringes for individual sites within the array for both base and MCM conditions (which is not sensitive to site-to-site variations in the phase shift), we find an average contrast reduction of 0.7(6)\%.  
The average phase shift, as well as site-to-site fluctuations about the average, can be mitigated either by calibrating their values and applying corrections to future operations, or using echo techniques -- for example swapping the qubit state populations between pairs of images such that the acquired phase shifts cancel instead of add.    The hiding light alone contributes negligible phase shift or loss of contrast.  Mid-circuit measurement of the $\ket{0}$ state has slightly worse performance, causing a contrast loss of 2.5 $\pm$ 1\%.  This is likely due to the presence of other imaging-laser frequency tones within the F = $3/2$ manifold (Appendix E).  Because of this factor (which is not fundamental), we focus primarily on mid-circuit measurements of $\ket{1}$ for the remainder of this work.

We confirm that hiding light applied to data sites causes minimal change to the imaging performance for ancilla sites.  Figure \ref{fig:fig2}c shows histograms of collected photons from the Ramsey sequence readout images, which do not have hiding light present (left), and from the ancilla sites during the mid-circuit measurement (right).  The data used here is averaged over the Ramsey fringe, and post-selected for occupied sites, which provides the even population distribution between bright and dark states.  The two histograms (which share identical imaging conditions, besides the presence of hiding light) are very similar, up to a several-photon greater peak separation for the mid-circuit measurement.  This could be attributed to an approximately 10~kHz red shift of the imaging transition on ancilla sites, which would imply a hiding beam intensity contrast at the part-per-ten-thousand level between data and ancilla sites.

In the two right-most columns of Table \ref{table:1}, we characterize loss and state leakage due to the mid-circuit measurement.  As expected, the loss probability per image is greatly reduced on hidden sites, as these atoms scatter very few photons on average.  For imaging of the $\ket{0}$ state, loss of non-hidden qubits is increased slightly (from 1.0(2) to 2.0(5)\%) when hiding light is on.  The reason for this increase is currently unknown, as we can constrain both heating and Raman scattering from residual hiding light to lower values based on the small change in average image brightness.

\begin{figure}[htb]
	\begin{center}
		\includegraphics[width=\columnwidth]{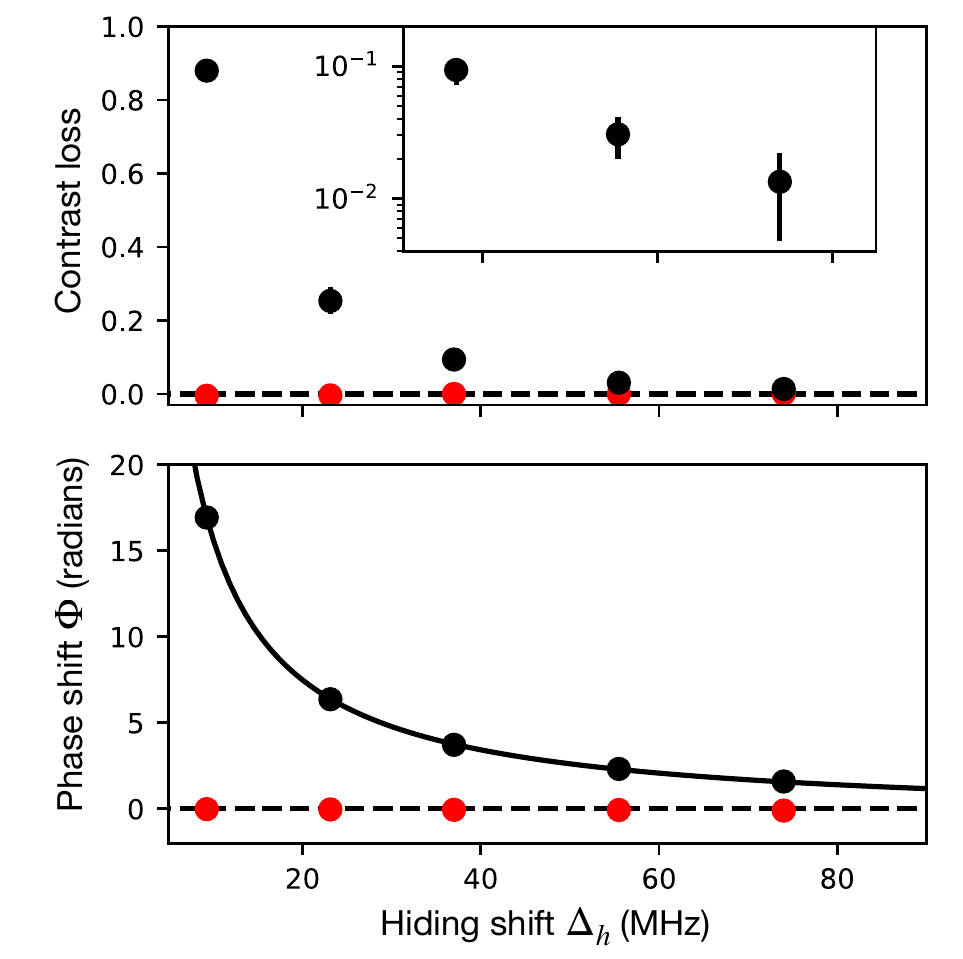}
		\caption{  Measured ensemble-averaged contrast loss (upper) and measured ensemble-averaged qubit phase shift $\Phi$ (lower) on data qubits due to hiding light alone (red points) and both hiding and imaging light combined (black points), versus the magnitude of hiding shift $\Delta_h$.  The inset in the upper panel shows the three points with greatest light-shift on a logarithmic scale.  The solid black line in the lower panel represents an inverse relationship between hiding shift and measurement-induced phase shift, with a constant offset to account for far-off-resonant tones.  Error bars correspond to those returned by a fit to the Ramsey fringe, and are smaller than markers when not visible.
		}
		\label{fig:fig3}
	\end{center}
\end{figure}

In figure~\ref{fig:fig3}, we characterize the performance of hiding versus the size of the light-shift $\Delta_h$.  The contrast maintained by data qubits improves with larger $\Delta_h$ over the range accessible given optical power constraints for the array size used here.  We expect that further improvements would be possible while maintaining scalability by using a hiding laser with a wavelength that has higher differential polarizability (nearer to a resonance between $^3$P$_1$ and a higher-energy state).  The phase shift imparted on the qubit (due to the off-resonant coupling of the qubit states to $^3$P$_1$) follows an approximate $1/\Delta_h$ scaling, with a coefficient of 164(4)rad MHz,  from which we extract a resonant saturation parameter for our imaging light of 1.2(4).  (We allow a constant offset when fitting the phase, to account for farther-detuned tones present on our imaging light.)

\section{correction of atom loss}\label{losscorr}
In typical quantum error correction protocols, the results of a mid-circuit measurement are used to apply conditional operations on qubits.  Further, ancilla qubits must be reinitialized after being measured.  As a proof-of-principle demonstration of these capabilities, we use the results of a mid-circuit measurement to correct for the occasional loss of ancilla qubits while maintaining coherence between data qubits.  

For this demonstration, we create a fully filled (above 98\% fill probability) 3 by 4 site sub-array by rearranging atoms from a 7 by 10 site array.  Individual atoms are moved from filled to empty sites using a single tweezer generated by crossed acousto-optical deflectors.  The sub-array is further subdivided into a checkerboard pattern of data and ancilla qubits.  Atoms remaining in the outer array form a reservoir used to refill the ancilla sites.  

\begin{figure}[htb]
	\begin{center}
		\includegraphics[width=\columnwidth]{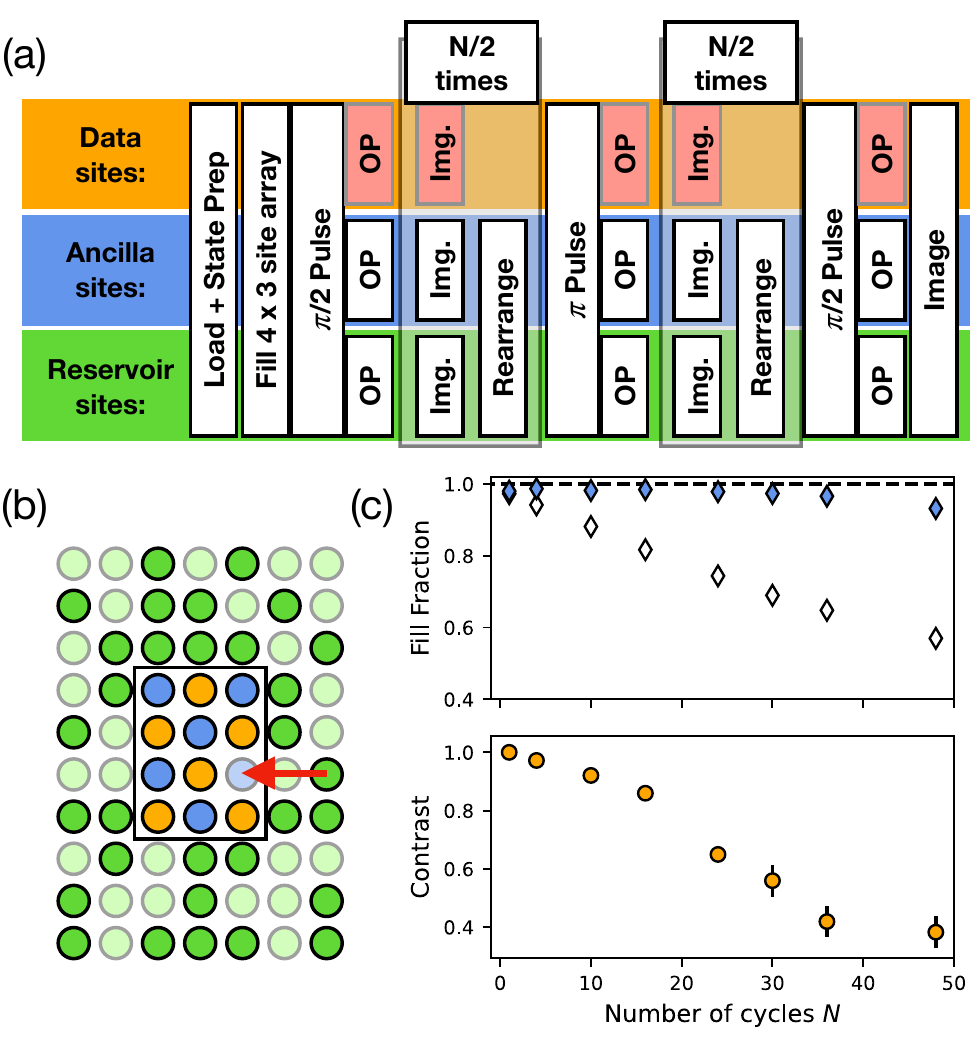}
		\caption{  Conditional reloading of ancilla qubits.  \textbf{(a)} Experimental sequence.  During repeated imaging and refilling of ancilla sites, data qubits are protected from decoherence by light-shifts from hiding beams, indicated by red boxes. \textbf{(b)} Data and ancilla sites (orange and blue, respecitively) are defined within a 3 by 4 sub-array surrounded by a reservoir region (green).  Ancilla sites identified as empty in mid-circuit images are refilled from occupied reservoir sites (red arrow). \textbf{(c)} Upper panel: ensemble-averaged fill fraction of ancilla sites with (blue diamonds) and without (white diamonds) conditional reloading. Lower panel: ensemble-averaged contrast on data sites relative to the $N=0$ case.   Up to 16 cycles, ancilla filling remains above 98\%, and contrast loss per cycle is 0.9(1)\% with conditional reloading.  Without reloading, ancilla filling drops by 1.1(1)\% per cycle. Error bars correspond to those returned by a fit to the Ramsey fringe, and are smaller than markers when not visible.
		}
		\label{fig:fig4}
	\end{center}
\end{figure}

As shown in detail in Fig.~\ref{fig:fig4}a, we embed $N$ repeated cycles of mid-circuit imaging of $\ket{1}$ and rearrangement operations on the ancilla qubits within a Ramsey sequence for the data qubits.  Optical pumping of the ancilla qubits after the global $\pi$ and $\pi/2$ pulses serves to reset their internal state to $\ket{1}$.
During optical pumping and imaging, the hiding light is applied to data sites in order to preserve coherence.  Optical pumping requires far fewer photons to be scattered than imaging does (we use 100~$\mu$s for OP compared to 5~ms for imaging, at similar power levels), and we have observed no impact on data qubit contrast from optical pumping.  We apply a single spin-echo pulse after $N/2$ cycles in order to reduce sensitivity to static differences in the qubit frequencies between sites, which are caused primarily by a gradient in the applied magnetic field.  On each cycle, ancilla sites identified as empty in the mid-circuit measurements are refilled with atoms from the reservoir.



By correcting for loss of ancilla atoms, we maintain an ancilla filling fraction of greater than 98\% out to 16 imaging and rearrangement cycles, beyond which it slowly drops, likely due to a lack of reservoir atoms on certain trials.  Without correction, the filling drops by 1.1(1)\% per imaging cycle for parameters used in this dataset.  Contrast loss per cycle out to 16 cycles is 0.9(1)\% for the data qubits, after which point it begins to deviate from an exponential decay, likely due to uncancelled coherent errors.  In these sequences, refilling of ancilla states takes place within 10-30~ms, with shorter times allotted for higher $N$ to keep the total Ramsey duration to approximately 600~ms.  The timing of the atomic movement has not yet been optimized for speed, and consists of 1~ms duration for each handoff to and from the moving tweezer, and 1~ms per site-spacing moved.  The remainder of the rearrangement time consists of camera readout and a buffer time.  Recent results on rearranging atoms within a similar static two-dimensional tweezer array indicate that with optimized parameters, movement of a single atom is possible within several hundred microseconds \cite{ebadi2021quantum}.

\begin{figure}[htb]
	\begin{center}
		\includegraphics[width=0.9 \columnwidth]{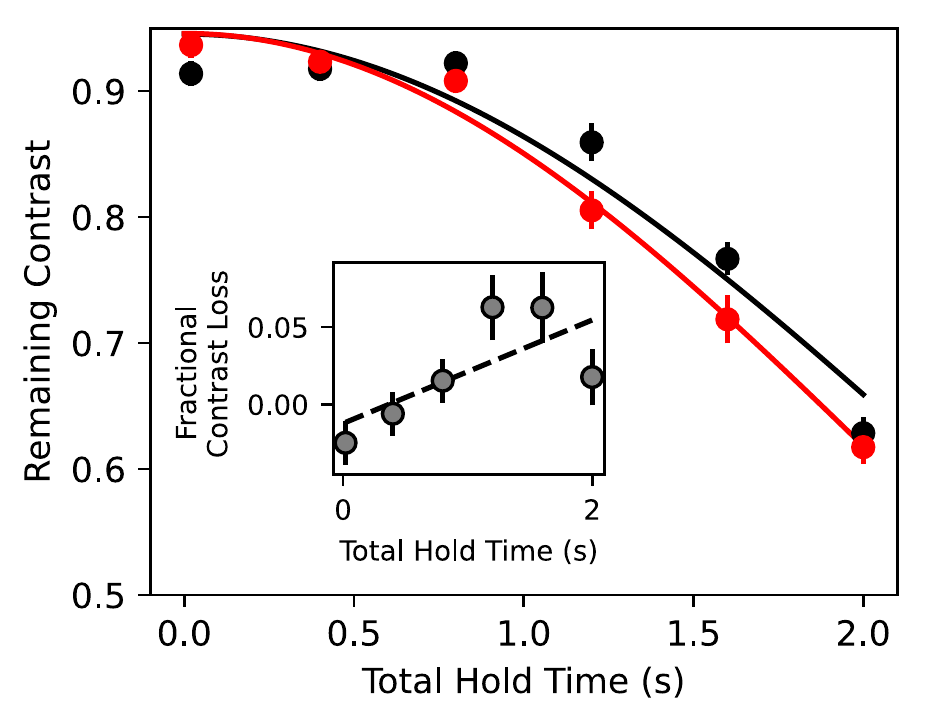}
		\caption{Coherence during MOT loading.  Main figure: remaining contrast after Ramsey sequence with (red points) and without (black points) simultaneous MOT operation.  A single $\pi$ pulse in the middle of the hold time is used to eliminate dephasing from static detuning errors.  Lines are Gaussian fits, to guide the eye.  Inset: relative contrast versus total hold time.  Error bars represent the standard deviation on the fitted contrast.  A linear fit indicates an additional contrast decay rate of rate of 0.03(2)/s in the presence of the MOT.  
		}
		\label{fig:fig5}
	\end{center}
\end{figure}
\section{Coherence during Magneto-optical trap Loading}\label{MOT}
In the long-run, it will be beneficial to not only reload ancilla qubits from a reservoir, but to replenish the reservoir while maintaining coherence among data qubits.  One challenge in doing so is to avoid qubit decoherence due to magnetic field gradients and scattered light from the magneto-optical trap (MOT) used to collect and cool atoms.  (This capability has recently been demonstrated in a two-species tweezer array \cite{singh2022mid}.) In particular, we use an initial MOT operating on the broad $^1$S$_0$ to $^1$P$_1$ transition near 399~nm.  This has a far greater risk of causing decoherence then the narrow-line MOT that follows (using the narrow $^1$S$_0$ to $^3$P$_1$ transition) both because the scattering rate in the broad-line MOT is higher and because Zeeman shifts in the high-field science region more dramatically suppress absorption on the narrow linewidth transition.  Our system utilizes a two-chamber design with static magnetic fields, and a 30~cm separation between the location of the MOT and the tweezer array.  This physical separation allows us to load atoms into the MOT while atoms in the science region remain coherent both by separating the coherent atoms from the spatial region where light is scattered and by allowing for the use of static magnetic fields.  We confirm this by running our standard MOT loading parameters during a Ramsey sequence on the qubits (with a single spin-echo pulse to eliminate the effects of static detuning errors between qubits).  
From a linear fit to additional contrast loss caused by the MOT, we extract an added decoherence rate of 0.03(2)/s (Fig.~\ref{fig:fig5}).  Our typical broad-line MOT loading lasts for 200~ms.  We attribute this residual decoherence to the absorption of photons scattered from atoms in the MOT.  The magnetic field gradients in the MOT region are static, and are constant in all experiments presented in this work.
The ability to transport and load atoms into our science array while retaining qubit coherence remains a future goal, but the risk of decoherence is limited because this process does not rely on scattering from a broad-linewidth transition. 


\section{outlook}\label{outlook}
The approach described in this work relies on a large ratio of transition frequency shifts -- Zeeman shifts for state-selectivity and light-shifts for site-selectivity -- to transition linewidth.  This makes it particularly well suited for alkaline-earth and lanthanide atoms that feature both narrow linewidth transitions suitable for imaging as well as broad transitions by which large light-shifts can be applied with moderate laser intensities, though it may be possible to adapt to narrow quadrupole transitions in Alkali atoms \cite{graham2023mid}.

Because of finite laser power availability, the desire for large hiding shifts creates tension with the desire to scale to large atom numbers.  In this work, we balance these desires to enable percent-level errors with 35 data sites, as such errors are commensurate with our current imaging performance.  The performance and scalability of hiding could be improved by using a hiding laser wavelength with larger differential polarizability for the imaging transitions.  This could be achieved either by tuning closer to the $^3$P$_1$ to $6s6d\,^3$D$_1$ transition used here (which would require a second laser if simultaneously attaining the magic condition for $^1$S$_0$ to $^3$P$_0$ is required), or operating the hiding array closer to a different transition such as $^3$P$_1$ to one of the higher-lying $^3$S$_1$ states \cite{zheng2022measurement}.  At increased hiding shifts, care will be required to limit the amount of residual hiding light on ancilla sites, as this can cause problematic shifts of the imaging light detuning and Raman scattering of near-detuned hiding light.  Additionally, technical improvements to our imaging system, such as reduction of background counts through better spectral filtering, or better photon collection efficiency, which we currently estimate to be approximately 4\%, could enable less imaging light to be applied to the atoms.  This would improve both the loss of contrast on data sites, as well as the imaging-induced loss on ancilla sites.

\section{Acknowledgements}
M.A.N.~and W.B.C~contributed equally.  We acknowledge helpful comments on the manuscript and system design from Jun Ye, and conversations with Monika Aidelsburger regarding the magic wavelength near 459~nm.  
During the preparation of this manuscript, we became aware of related work using a similar state-selective imaging technique \cite{huie2023repetitive}.  Submission of this manuscript was coordinated with related work from Adam Kaufman's group \cite{lis2023mid}.  
\section{Appendix A: Experimental apparatus and sequence}

Our experimental system consists of two main vacuum regions -- the ``MOT chamber" and the ``science chamber" -- connected by a differential pumping tube.  Atoms are loaded from a pre-cooled atomic beam into a two-stage magneto-optical trap (formed using the 399~nm $^1$P$_1$ transition followed by the 556~nm $^3$P$_1$ narrow-line transition) in the MOT chamber.  Atoms are then loaded into an optical lattice formed using 532~nm light and transported vertically by 30~cm into the science chamber.  In order to achieve a deep lattice in a power-efficient manner, the waists of the transport beams are translated synchronously with the optical lattice by moving the position of two focusing lenses, one for each of the two counterpropagating beams that form the lattice.  Alignment between the two beams is actively maintained using closed-loop piezo-electric steering mirrors.  Atoms are transferred into the optical tweezer array by overlapping the atoms with the array, ramping up the power in the tweezers, and then ramping down the transport lattice.  This leads to a typical occupancy of several atoms per tweezer.  No dissipation is applied to transfer atoms from the transport lattice into the tweezers.  To our knowledge, this is the first example of loading thermal atoms into an array of micron-scale tweezers without applying dissipative cooling.  

Our two-chamber design allows us to work with temporally static magnetic fields.  No magnetic fields are varied during any of our experimental sequences, which both avoids time-delays associated with switching, and allows us to simultaneously maintain a magnetic field gradient for MOT formation and a large and uniform bias field in our science region.  Our two-stage MOT operates at a constant field gradient of approximately 18~Gauss/cm in the strong direction.

After atoms are loaded into tweezers, we apply light with the same parameters as used for imaging of the $m_f = 1/2$ qubit state to induce light-assisted collisions and project to a single atom per tweezer.  During this time, a second tone that addresses the $^1$S$_0$, $m_f = -1/2$ to $^3$P$_1$, $m_f = 1/2$ transition is applied to transfer all atoms to the $m_f = 1/2$ state.  

\section{Appendix B: measurement of errors}
This section describes the experimental sequences and analyses used to determine the sources of measurement error presented in Table \ref{table:1} of the main text, as well as estimates for their primary physical origins.  All values represent ensemble averages over a 7 by 10 site array that is stochastically loaded (approximately 50\% fill fraction).  For the ``base" condition, where no hiding light is applied, results from all sites are averaged together.  For the mid-circuit measurement results (final two data columns), the 7 by 10 site array is subdivided into a checkerboard of data and ancilla sites, with hiding light at the default conditions described in the main text ($\Delta_h = 74$~MHz) applied to the data qubits.  Data and ancilla sites are then separately averaged together to obtain the values reported in the table.  

Overlap errors (first and second data rows) are estimated by performing repeated images of stochastically loaded sites with atoms prepared in the state to be imaged using optical pumping.  Photons are recorded on a low-noise camera, and the results are summed over rectangular regions of interest centered on each site (each region is 3 by 3 pixels, corresponding to an approximately 1.5~$\mu$m square in the atomic plane).  For display purposes (Fig.~\ref{fig:fig2} c), a constant offset corresponding to the typical level of background counts (dominated by the arbitrary offset reported by the camera) is subtracted, and the results are scaled by the specified sensitivity of the camera.  We note that these offsets and scalings do not impact our quantitative estimate of overlap error.  

To estimate the overlap error, we fit a sum of two Gaussian profiles to the histogram formed from the images of stochastically loaded sites.  We then define a threshold value to distinguish between bright and dark sites.  For simplicity, we choose the threshold that minimizes the error corresponding to the summed weight of the tails on the incorrect side of the threshold for two equal-area Gaussians.  If certain types of error (false positives or negatives) are more damaging for a given use-case, a different criteria for determining the threshold could be applied.  For the datasets presented in Fig.~\ref{fig:fig2}c, the error determined in this manner is below 0.02\%. In Table \ref{table:1}, we report a more conservative minimum error of 0.1(1)\% for ancilla during MCM to account for the possibility of day-to-day drifts in parameters, and for possible deviations of the true distribution from a double-Gaussian profile.  The values reported for the base condition represent specific measured values, with the error representing typical drifts.   Because our overlap error is an order of magnitude smaller than the error associated with loss, it may be possible to improve total performance by using less intense imaging light or a shorter image.  However, because the overlap error worsens rapidly with lower counts while the loss error improves only linearly, we find the current conditions to be a favorable balance of performance and robustness to fluctuating scattering rates.  

To measure loss of the imaged state in the base condition, (in data rows 3, 7 of Table \ref{table:1}) we perform twenty subsequent images (performed with our default parameters and duration) of stochastically loaded atoms.  We fit an exponential decay to the array-averaged bright fraction, and report the decay fraction per image as loss.  The uncertainties reported on the loss rate for the imaged state represent the uncertainty of the parameter estimation from the exponential fit.   
The loss estimated in this manner is a combination of actual imaging loss and any loss that occurs in the time between images.  We include 21~ms between each image for camera readout, which leads to a total duration of 26~ms per image.  At our measured typical vacuum lifetime of 30~seconds, background gas collisions contribute to 0.001 loss per image to the measured value, though the majority of this is not during the image itself.  Further mechanisms of loss (whose estimates are coupled with those of spin flips) are described in Appendix C.  

The probabilities of spin flips, loss of the non-imaged state, and loss of data qubits are much lower than the loss of the imaged state on non-hidden sites, so we use much longer ``dummy" images to induce the error (rows 4-6, 8-10 and final two columns of rows 3, 7 of Table \ref{table:1}). First, we prepare the atoms in the desired state using optical pumping, and then perform a pre-image of that state in order to determine site occupancy.  Only data from sites determined to be occupied is included in further analysis.  We then perform the dummy image of one of the two qubit states using identical parameters to our default images, except for the duration.  We use 200~ms dummy images for the base condition, and 50~ms dummy images for the MCM condition where certain error rates are higher, though this difference is not strictly necessary for the loss rates observed with the default MCM parameters.  Finally, we perform a final image of one of the two qubit states (either the same state that was prepared to determine loss or the opposite state to determine spin flips) and compute the fractional loss or probability of atoms appearing in the imaged state, scaled to the ratio of default image duration (5~ms) to dummy image duration.  We note that loss, overlap, and spin-flip errors associated with the pre-image and final image are also scaled down by this ratio, and we apply no corrections for measurement fidelity.  The uncertainty reported for these measurements represent the Wilson score interval, a method for estimating the uncertainty of parameter estimation for a binomial distribution \cite{wilson1927}.  While the Wilson score interval is asymmetric, we display only the larger of the two error directions in Table \ref{table:1} for visual clarity.

\section{Appendix c: estimates of error sources}
This section describes estimates for the magnitude of processes that cause the error terms presented in Table \ref{table:1}.  

Raman scattering of population out of the upper imaging state ($^3$P$_1$, $m_f = \pm 3/2$) due to trap or hiding light can lead to atom loss from the imaged state as well as spin flips.  Atoms decaying to the long-lived $^3$P$_0$ and $^3$P$_2$ states are not repumped to the qubit manifold, and barring subsequent scattering events (negligible in our short experimental sequences), appear as loss.  Population scattered into $^3$P$_1$ states other than the desired stretched state may decay into the spin state not being measured, contributing a finite probability of spin flips.  

We calculate the expected rates of Raman scattering using the Kramers-Heisenberg formula as described in Ref.~\cite{doerscher2019lattice}. At our trap depths of 350~$\mu$K, we expect that the Raman scattering rate for an atom in $^3$P$_1$, $m_f = \pm 3/2$ into $^3$P$_0$ or $^3$P$_2$ to be 0.6/s and 1.73/s respectively.  It is convenient to express these rates in units of the $^3$P$_1$ decay rate of $1.14 \times 10^6$/s, to define a branching ratio $R_i$ for the decay channel $i$.  Based on our estimated collection efficiency $\eta$ typical number of collected photons $N$ we can extract an estimate for the single-image error rate associated with that channel: $N R_i/\eta$.   We predict the branching ratios into $^3$P$_0$ and $^3$P$_2$ to be $5 \times 10^{-7}$ and $1.7 \times 10^{-6}$ respectively. Using typical values of $N = 30$ and $\eta = .04(1)$,  the expected combined effective loss rate from these channels is then $1.5(5) \times 10^{-3}$ in a typical image, where the reported uncertainty corresponds to imperfect knowledge of our collection efficiency.  As pointed out in \cite{ma2023high}, because the 483~nm light used for trapping is sufficiently energetic to photo-ionize $^3$P$_1$ atoms via the absorption of two photons, photo-ionization may contribute additional loss from $^3$P$_1$.  Further, in order to improve the speed of our imaging, we operate at a photon scattering rate that may contribute a small amount of additional loss due to heating.  

Atoms that Raman scatter into other states within the $^3$P$_1$ manifold can either decay into the original qubit state, or decay into the opposite qubit state causing a spin flip.  By summing the probabilities of Raman scattering into the different $^3$P$_1$ states, weighted by their probabilities of causing a spin flip upon decay, we calculate the expected spin flip probability for a typical image to be $9(3) \times 10^{-5}$.  

Because our hiding light is near-detuned from the $^3$P$_1$ to $6s6d\,^3$D$_1$ transition, we expect Raman scattering rates for atoms populating $^3$P$_1$ to be roughly two orders of magnitude higher than for our trapping light (we use similar intensities for hiding light and trapping light).  However, the effect of this Raman scattering is negligible because the hidden atoms do not populate $^3$P$_1$ (as we confirm from the percent-level contrast loss: 0.01 photons are typically scattered for hidden atoms during an image compared to approximately 750 for non-hidden atoms), and leakage hiding light onto non-hidden atoms is bounded to an intensity four orders of magnitude lower than the trapping light.

In addition to Raman scattering out of $^3$P$_1$ due to hiding or trapping light, spin flips can be caused by off-resonant scattering of imaging light from the $^3$P$_1$, $m_f =  \pm 1/2$ states.  The dominant polarization component used for imaging can cause scattering from the non-imaged state into the imaged state, while polarization impurities may cause scattering from the imaged state to the non-imaged state.  These processes are strongly suppressed by our large applied magnetic field and correspondingly large detunings, with contributions scaling as the inverse of the square of detuning.    Because of the multiple tones of imaging light present in the $F = 3/2$ manifold (Fig.~\ref{fig:detunings}), lack of precise knowledge of the modulation depth applied to generate the imaging sidebands, and uncertainty in the precise detuning of the imaging light, a precise estimate of the scattering rate is difficult.  However, we can obtain an estimate for what is likely the largest channel for spin flips due to scattering: $\ket{1} \rightarrow \ket{0}$ transitions due to the upper first order sideband during imaging of $\ket{0}$ (the lower first-order sideband provides the desired imaging light).  This sideband is approximately 200~MHz detuned from the $\ket{1}$ to $F = 3/2$ $m_f = -1/2$ transition, with the correct polarization to drive the transition.  Based on our estimated imaging saturation parameter of 1.2(4) (main text), and accounting for Clebsch-Gordan coefficients, we estimate the spin flip probability associated with this mechanism to be $1.5(5) \times 10^{-4}$ during an image, consistent with our observations.  Other scattering channels that might cause spin flips are either farther detuned, involve higher-order (weaker) sidebands, and/or are driven only by polarization impurities, and so are expected to be smaller.  

In principle, imaging light could drive two-photon Raman transitions between qubit states.  However, due to the large detuning from intermediate states and the fact that our magnetic field is oriented along the propagation direction of the imaging light (greatly suppressing the strength of the $\pi$ component required to drive the transition), we estimate the two-photon Rabi frequency to be on the Hz scale.  Because this is much smaller than qubit frequency (388.9~kHz), spin-flips associated with this mechanism are expected to be negligible.

The two primary mechanisms by which mid-circuit measurement causes loss of contrast on data qubits are the residual scattering of imaging light and spatially or temporally varying phase shifts.  The impact of spatially varying phase shifts are discussed in appendix D.  Because the light-shifts applied for hiding are finite, we expect some residual scattering to be present in the data qubits on the transition used to image.  For our default conditions ($\Delta_h = 74$~MHz and a saturation parameter of 1.2(4)), we expect a scattering probability (and associated loss of contrast) of 0.005(2) from the imaging transition.  The presence of other tones within the $^3$P$_1$, $F =  3/2$ manifold (Fig.\ref{fig:detunings}) may cause additional smaller loss of contrast, in particular when measuring state $\ket{0}$.

\section{Appendix D: Single-site contrast and phase shifts}
\begin{figure*}[htb]
	\begin{center}
		\includegraphics[width=1.5\columnwidth]{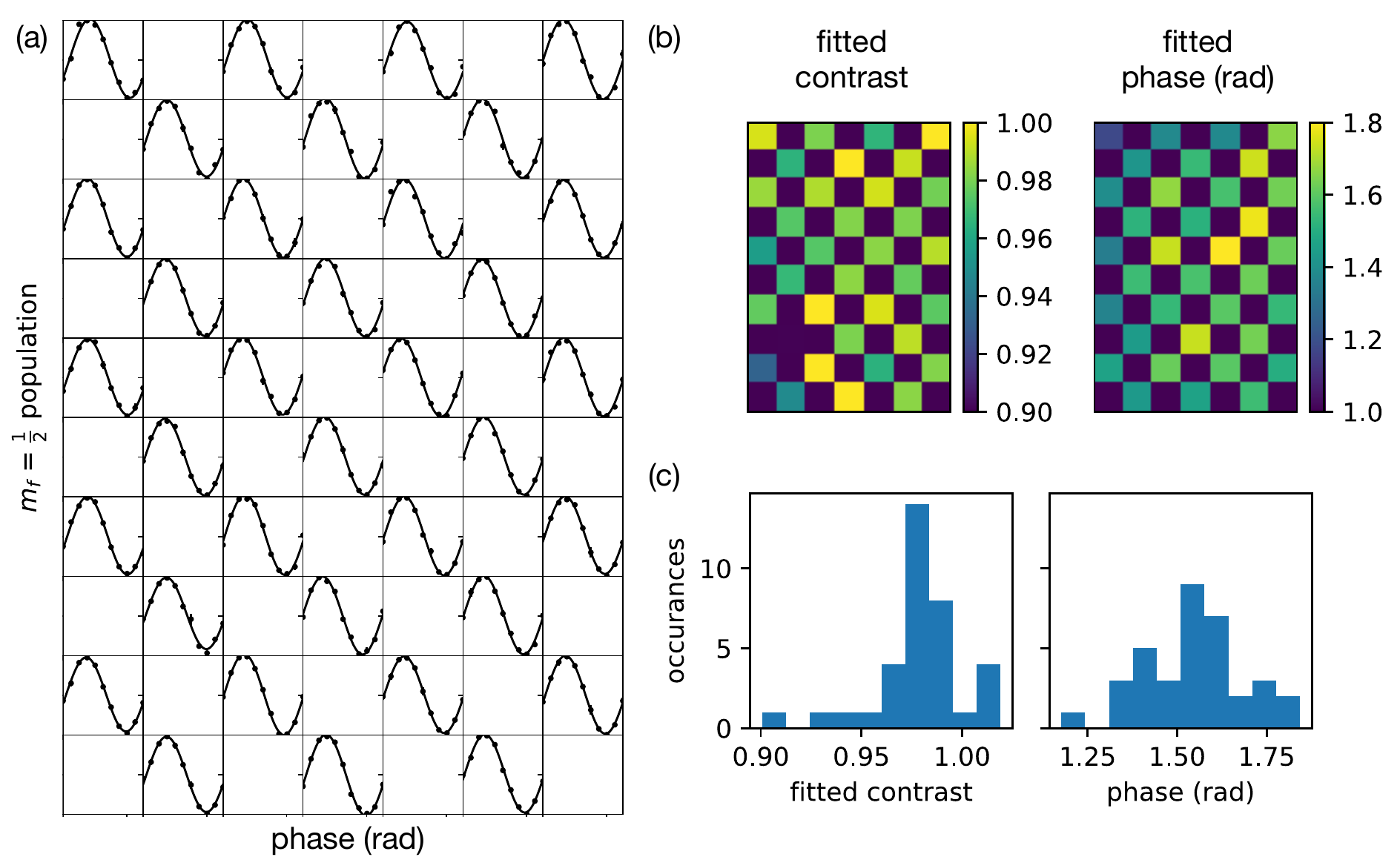}
		\caption{\textbf{(a)} Site-resolved data-qubit Ramsey contrast fringes in the presence of MCM on ancilla qubits (divided in a checkerboard pattern), with individual cosine-fits.  Where not visible, error bars are smaller than markers.  Ancilla sites are omitted for clarity.  Axis ranges for each subplot are 0 to 1 vertically and 0 to $2\pi$ horizontally.  \textbf{(b)}  Fitted amplitudes and phases for individual sites, displayed according to their position in the array.  \textbf{(c)} Histograms of contrast and phase from individual sites.  (The overall offset to phase is arbitrary.)}
		\label{fig:singles}
	\end{center}
\end{figure*}

Unless stated otherwise, all quantitative data presented in the main text is averaged over all atoms within the array of a certain type (data, ancilla, reservoir) prior to performing any fits.  In figure~\ref{fig:singles}, we present a site-by-site analysis of the MCM data from figure \ref{fig:fig2}b (orange curve).  Though the single-site statistics are worse than the array-averaged ones, we can still fit the Ramsey fringes from individual sites (Fig.~\ref{fig:singles}a).  For visual clarity, we plot only data sites here.  Ancilla sites are fully decohered at the single-site level as well as when array-averaged.  We show the fitted amplitudes and phases across the array in figure~\ref{fig:singles}~b,c.  Averaging the individually fitted amplitudes across the array gives a contrast of 98.7(4)\% for the base condition (no measurement or hiding light) and 98.0(4)\% with MCM, corresponding to a 0.7(6)\% reduction in contrast, with the error defined by the standard deviation of the mean of the individually fitted contrasts.  This contrast loss is consistent with our estimate provided above for residual scattering of imaging light, and can be compared to the ensemble-averaged contrast reduction of 1.3(8)\% (averaging data prior to fitting).  The former value is insensitive to site-to-site variations in phase shifts, while the latter is sensitive to these shifts.  The fitted phases vary with a standard deviation of 0.14 radians (compared to 0.05 radians for the base condition), which would contribute a 1.0\% loss of contrast.

\section{Appendix E: Laser tones within the  $^3$P$_1$, $F = 3/2$ manifold}
\begin{figure}[htb]
	\begin{center}
		\includegraphics[width=\columnwidth]{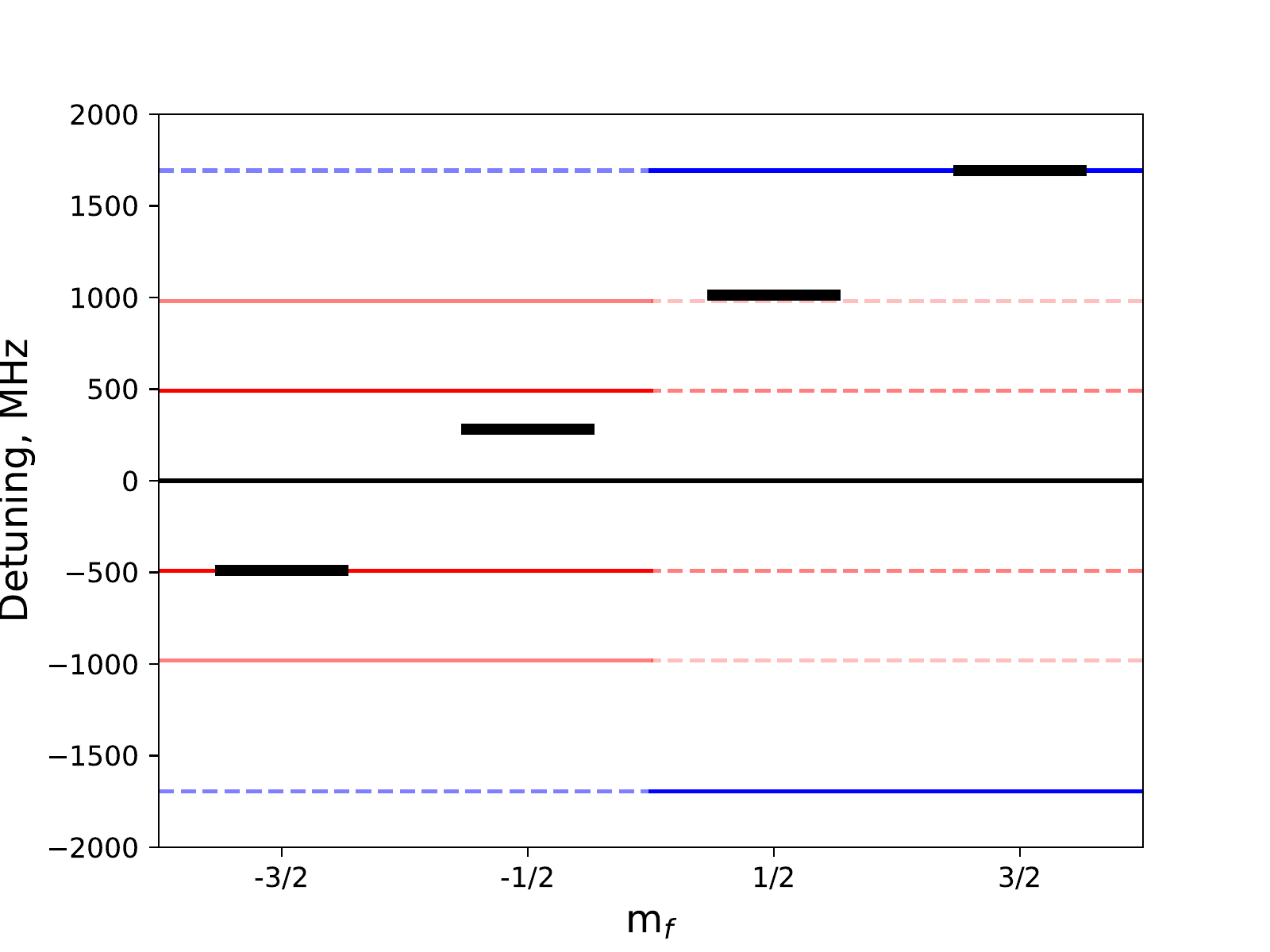}

		\caption{  Frequencies of imaging laser tones within the $^3$P$_1$ $F = 3/2$ manifold.  Atomic states are represented by thick black lines.  The carrier tone is the thin black line.  EOM sideband tones corresponding to imaging of the $m_f = -1/2$ ($\ket{0}$) state and the $m_f = 1/2$ ($\ket{1}$) states are represented in red and blue, respectively.  Solid lines indicated where states may couple to the qubit states via the dominant polarization component, and dashed lines indicate where coupling occurs through impure polarization.  Second order sidebands are indicated by lower color saturation than first order.   
		}
		\label{fig:detunings}
	\end{center}
\end{figure}

For this work, we address the $^1$S$_0$ to $^3$P$_1$ transitions using light derived from a single laser, incident along two counter-propagating paths.  Each path has a fiber acousto-optic modulator, which can provide rapid power switching and high extinction.  Additionally, each path has a fiber-coupled electro-optic modulator (EOM), which can be used to apply sidebands at frequencies up to several GHz. 
Because the same laser is also used to form the MOT, the carrier tone of the imaging beams is within the $^3$P$_1$, $F = 3/2$ manifold (Fig.~\ref{fig:detunings}), detuned 280~MHz to the red of the $^3$P$_1$, $m_f = -1/2$ state.  When imaging the $m_f = -1/2$ qubit state (red lines), this also places other EOM sidebands within the $^3$P$_1$, $F = 3/2$ manifold, including a second order sideband approximately 34~MHz to the red of the $^3$P$_1$ $m_f = 1/2$
state.  Imperfect polarization on the imaging beam may cause unwanted scattering from this sideband.  We attribute the greater reduction in contrast when imaging the $m_f = -1/2$ qubit state to the presence of these additional tones.  This limitation could be overcome by using a separate laser for the imaging light, or by using high-frequency or multi-passed acousto-optical deflectors.   

\section{Appendix F: Determination of $^3$P$_0$ magic wavelength}
To determine the magic wavelength for the clock transition, we transfer atoms from the $^3$P$_1$-magic traps used for imaging into the clock-magic tweezers where we perform $^3$P$_0$ spectroscopy with a global clock beam, and back to the $^3$P$_1$-magic traps for readout.  For spectroscopy, we use a Ramsey sequence with a single $\pi$ pulse in the center (30~ms before and after the $\pi$ pulse), and vary the intensity of the trapping light before the $\pi$ pulse.  For non-magic wavelengths, this leads to a sinusoidal variation in population versus trap power imbalance, and we tune the trap wavelengths to remove these variations.  Compared to traditional Rabi spectroscopy, this method provides insensitivity to drifts in the clock laser frequency.  

We find an optimal wavelength for the trapping light of 459.5960(5)~nm, measured against a wavemeter calibrated by relative to the $^3$P$_1$ transition.  The reported error is dominated by potential drifts of the wavemeter.  

\bibliography{bib}
\end{document}